# The Perfect Way to Manage Spectrum

William Webb, *Fellow, IEEE*, Arturas Medeisis, *Senior Member, IEEE*, and Leo Fulvio Minervini

*Abstract*—This article discusses the key principles of radio spectrum management with a focus on spectrum allocation and access. We show the current regime's inherent rigidity and constrained possibilities for introducing new radiocommunication services and applications. The article proposes how the governments and spectrum users could cooperate in taking spectrum management to a qualitatively new level, characterized by light touch regulation and flexible use. This could be achieved through broader introduction of emerging practices such as Spectrum Usage Rights, liberalised spectrum trading, and full shared spectrum access. We conclude by presenting a vision for a "perfect" spectrum management arrangement and future research directions.

## I. INTRODUCTION

A key reason for managing spectrum is to avoid interference between different users – if all users were able to access the spectrum without any planning and yet without interfering with each other, then there would be no need for spectrum management. About a century ago, there was a short interlude of open access, in which users generally coexisted without interference. But then radio services began to grow dramatically and governments quickly adopted "command-and-control" methods to manage spectrum [1].

To avoid interference, the usual arrangement is to give spectrum users administrative licences which set out in some form their "rights". It is important that neighbours both in geography and in spectrum are given compatible licences – i.e., that they do not cause excessive interference to each other. In addition, a few parts of spectrum are open for unlicensed use, known as spectrum commons.

Besides ensuring technical compatibility, the spectrum should be managed so that it ensures equitable access to various players and delivers the greatest possible contribution to GDP as well as other desirable outcomes like enabling various functions of the state and social development. This also means that "perfect" spectrum management cannot occur without some form of government oversight [1], [2].

The objective of this article is to motivate both regulators and industry thought leaders in considering new and improved ways for managing radio spectrum more efficiently, and to encourage further research in this important field. In our proposed vision, the "perfect" spectrum management should be as light touch as possible, with centralised control retained only for public spectrum uses, while the bands allowed for commercial use to be increasingly governed through decentralised and self-regulating approaches.

## II. WHAT IS PERFECT WAY TO MANAGE SPECTRUM?

The term "perfect" used in this article is meant to be inspirational rather than represent some formal definition of technical or economic efficiency. However, it is now a common belief that the economic efficiency of spectrum use should be the foremost objective of good spectrum governance.

To deliver the maximum economic efficiency each "piece" of spectrum needs to be used for the radiocommunication service that has the greatest value and within that service in the most technically efficient manner over time. This implies there need to be changes in the allocation of spectrum across uses (including service rules) and across users whenever this could lead to an increase in the value that society places on spectrum. Importantly, this definition takes into account the value that society may place on spectrum allocations and assignments that are not market driven [1].

It is however important to note, that there may be also public interest requirements, that a pure market regime would not meet, and those result in some spectrum being set aside for socially valuable uses rather than those that are only economically efficient – this will tend to be a political decision and will normally require direct intervention.

Those considerations also hint at economic efficiency of public-sector spectrum use. In principle, spectrum held by the public sector should be subject to perfect management. However, improving efficiency in the public sector involves greater complexity than in the private sector as the public sector has more opaque incentives. In particular, the competitive forces that typically promote a firm's desire to minimize costs, inter alia the spectrum acquisition cost, are often absent in cases of spectrum use by governmental users [1]. As regards the applicability of trading to public sector spectrum, the government spectrum users face uncertainty in retaining the surplus associated with trading some of its spectrum as it may be clawed back in any budgetary process.

Few countries, e.g. the UK, have been engaged in the reform of public-sector spectrum management [1], but use of spectrum by the public sector remains largely inefficient.

William Webb is with Access Partnership, London SW1E 6QT, UK (e-mail: william.webb@accesspartnership.com).

Arturas Medeisis is with the Department of Computer Science and Communications Technologies, Vilnius Gediminas Technical University, LT-10223 Vilnius, Lithuania (e-mail: arturas.medeisis@vilniustech.lt).

Leo Fulvio Minervini is with the Department of Economics and Law, University of Macerata, 62100 Macerata, Italy (e-mail: leofulvio.minervini@unimc.it).

Overall, the perfect spectrum management should be the lightest touch possible. This is because regulatory activity:

- has a direct cost associated with it, which is often met by the users of spectrum;
- imposes bureaucratic burdens on users;
- slows down change arising from technical or other advances, innovations, or new concepts;
- is prone to regulatory capture, where incumbents can influence outcomes in a way that preserves their interests;
- is generally considered to be less likely to deliver the perfect outcome than the market since the regulator cannot have as much insight as the combination of all interested entities.

Hence, the perfect spectrum management regime would be one where a regulator was not required (once it had established a framework for management), or only needed to enable a socially desired outcome. In practice there may be some areas where this is just not possible, especially those that require international coordination such as satellite and aeronautical usage, and where market forces cannot operate effectively, such as for unlicensed, amateur radio, or passive spectrum uses.

As well as minimising the actions of the regulator, a perfect regime would decrease bureaucracy by reducing the need for licences. Approaches that do not require a licence, such as use of spectrum commons, or that only require minimal interaction, such as database-driven light licensing or Dynamic Spectrum Access (DSA), should be preferred.

### III. HOW GOOD IS THE CURRENT REGIME?

Current spectrum management falls far short of this ideal. Generally, once a piece of spectrum has been allocated for a specific use (e.g., broadcasting) it is not possible for anyone, except the regulator, to change that.

Regulatory change of use, often involving complex inter-governmental negotiations and consensus building in various study groups and conferences, is very slow – taking at least 5 – 10 years to reach a decision, clear the spectrum and then award it. Even large countries that try short-cutting this process and reallocate spectrum unilaterally, end up encountering severe obstacles and uncertain outcomes [3].

Each country has its own spectrum management agency – a National Regulatory Authority (NRA). The NRAs cooperate internationally on standardization and harmonization of spectrum use within the framework of the International Telecommunication Union (ITU) and several regional organizations (CEPT in Europe, CITEL in Americas, etc.) [4].

NRAs and ITU dedicate large portion of their spectrum management activities to advance the strategic planning of future uses of various frequency bands and setting of respective spectrum access policies and licensing principles. The key term establishing the fundamental conditions for spectrum access is the service allocation, which links a specific frequency band to the type of radiocommunication service allowed in that band. The ITU Radio Regulations (RR) – an international treaty document – defines more than 40 types of radiocommunication services used to describe service allocation in a particular band. Fig. 1 offers a glimpse on some of the top layer RR radiocommunication service categories.

The Article 5 of the ITU RR, spanning some 150 pages, enumerates such service allocations for rather arbitrarily partitioned bands across the radio spectrum from 8.3 kHz to 275 GHz. Once approved by an ITU World Radiocommunications Conference (WRC), the RR service allocations are then transposed by NRAs into what is known as a National Table of Frequency Allocations.

Fig. 2 depicts the two principal paradigms of accessing radio spectrum and the examples of their implementation through various forms of access mechanisms. They establish principal conditions for spectrum access and may convey certain usage rights. Those could be then enacted in practice by means of the frequency assignment process, be it some form of administrative assignment or market-based mechanism, or indeed some form of self-managed spectrum access [4].

The current approach to service allocation and establishing conditions for spectrum access is far from being truly technologically neutral. It is too much fixated on the notion of completely avoiding interference, by means of effectively isolating each licensed radio system in time-area-frequency domains.

| | |
|---|---|
| Aeronautical Mobile | Aeronautical Mobile-Satellite |
| Aeronautical Radionavigation | Aero Radionavigation-Satellite |
| Amateur | Amateur-Satellite |
| Broadcasting | Broadcasting-Satellite |
| Fixed | Fixed-Satellite |
| Land Mobile | Land Mobile-Satellite |
| Maritime Mobile | Maritime Mobile-Satellite |
| Maritime Radionavigation | Maritime Radionav-Satellite |
| Meteorological Aids | Meteorological-Satellite |
| Mobile | Mobile-Satellite |
| Radiodetermination | Radiodetermination-Satellite |
| Radiolocation | Radiolocation-Satellite |
| Radionavigation | Radionavigation-Satellite |
| Standard Frequency and Time | Std Freq & Time-Satellite |
| Radio Astronomy | Earth Exploration-Satellite |
| Ship Movement | Inter-Satellite |
| Port Operation | Space Research |

**Fig. 1.** A sample of radiocommunication services to which various frequency bands are allocated in ITU RR.

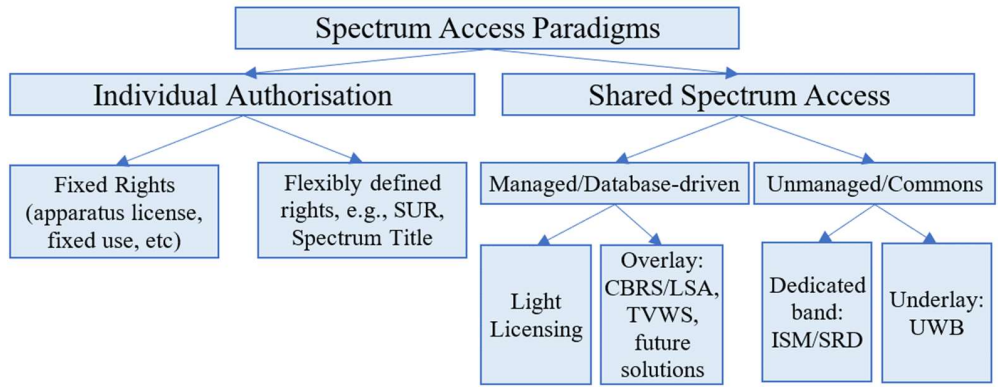

**Fig. 2.** Spectrum Access Paradigms.

In simple terms, this is done by ensuring that the interfering signals do not exceed a certain minimal Signal-to-Noise Ratio (SNR) threshold at the edge of service area of victim system, as illustrated in Fig. 3. Note that the time domain is not considered here, which corresponds to most frequent assumption of full-time operation.

Such approach predicates that the NRAs and ITU study groups that prepare WRC decisions, need to fix *a priori* assumptions on the type of radiocommunication systems that may be allowed to operate in given band. That includes such technical parameters as maximum transmit power, receiver sensitivity and target SNR criteria, as well as channelling and duplex arrangements inside the band – i.e., use of either Frequency Division Duplex or Time Division Duplex.

The resulting spectrum allocation decisions, and the subsequent frequency assignment process, institute explicit static separation distances in frequency (guard bands) and area (minimum separation distances) domains. The spectrum use efficiency becomes an afterthought rather than the primary target, and, more often than not, is suffering for that. This may be well seen in Fig. 3, which shows the significant areas becoming unavailable for operation.

This inherent wastefulness of the traditional spectrum management approach needs to change, to make the most efficient use of spectrum and provide adequate access options for new uses and broadband applications [5], [6].

When deciding on phasing out legacy systems, the NRAs may justifiably need to make technical assumptions on the likely future uses of the band. This is a necessary starting foundation to assess the spectrum co-existence solution and carry out frequency assignment to new users. But once the spectrum has been (re-)assigned, it should not be seen as locking out any future change of use.

Given the above discussed practices, the current spectrum management regime is decidedly not light touch. Critically for business and innovation, there is a substantial overhead for wireless industry players, associated with lobbying the regulator, responding to consultations, and, in some cases, undertaking legal challenge.

Current spectrum management also rarely allows the more efficient forms of sharing, so that some bands are inefficiently used, especially those held by governments for occasional uses.

Hence, current regimes are in no way perfect. Even though NRAs do try some innovations and pioneer new thinking, still until today they retain their role as the main authority in all questions of spectrum access and usage [2], [7], [8].

## IV. WHAT IS NEEDED?

It is important to start by admitting that the existing spectrum management regime needs to evolve [4], [9], [10]. Enhancing spectrum use efficiency, as well as flexibility and certainty in

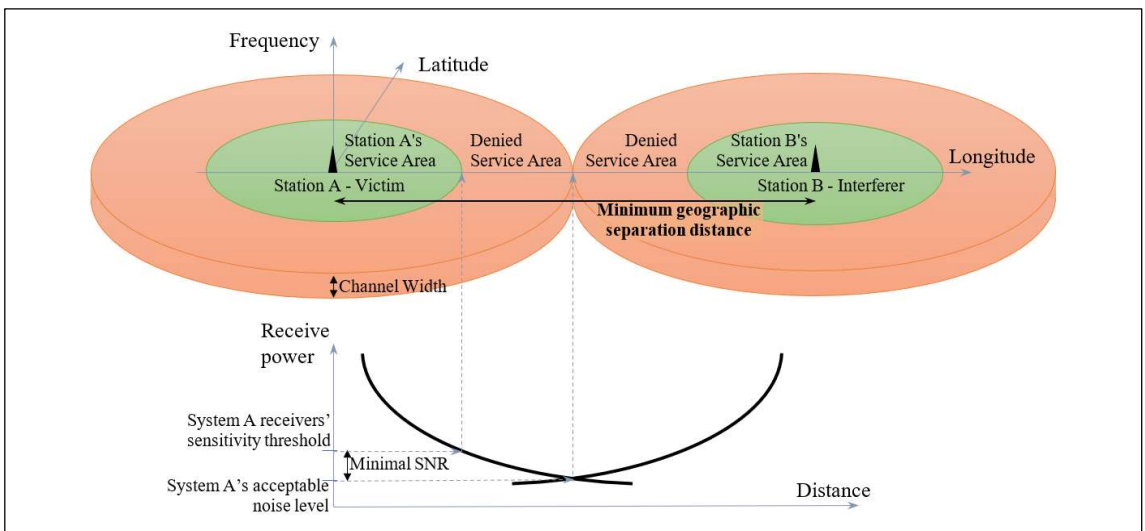

**Fig. 3.** SNR-based delineation of spectrum use space.

spectrum access should become the priority. To deliver on these objectives, the main requirements are:

1) Perpetual licences with full service neutrality, so any spectrum can be used for any application and without concerns as to time constraints for recouping any investments [2], [5].
2) Full shared access so that any unused spectrum can be exploited by others [9], [10].
3) Full trading in all its various aspects (e.g., trading, leasing, sub-division of resource, trading of boundary parameters) to allow the market to work effectively [1], [11].
4) Applying regulatory intervention only when needed (e.g., for socially desirable outcomes) and in a way that has the lowest impact on commercial use [2], [5], [7].

These are discussed in turn below.

### A. Service neutrality

The reason for restricting the type of use of a piece of spectrum is to avoid interference occurring, mostly to users in neighbouring frequency bands, but also potentially across geographical border areas. If usage changes (e.g., from broadcasting to cellular) then, under current licensing approaches that tend to restrict transmitter power, it is quite likely that interference will change. There have been cases (e.g., US: Nextel) where this has resulted in the need of costly intervention to resolve.

There is a simple solution to this – change the licensing approach to restrict total interference caused rather than power transmitted. An approach that enables this has already been explored in detail and used in practice – the Spectrum Usage Rights (SURs) developed by UK's Ofcom [12].

The SUR concept may be seen as a modern evolution of the Nobel prize winning studies of Ronald Coase, developed some sixty years ago. Coase advocated the partial spectrum privatisation with associated frequency zoning, like a land use plan, to allow spectrum use under certain border conditions set out by the state to control interference [13]. Following Coase's approach, the spectrum that is meant for commercial exploitation should be turned into property and freely bought and sold in the marketplace. Note in that regard the caveat regarding spectrum for public uses, as discussed in Section II.

**Fig. 4.** SUR-based delineation of Spectrum-Area use space.

Trading within the framework of a system of property rights will provide the right economic incentives to take interference into account and thus lead to an efficient use of spectrum, provided that the property rights are fully and precisely defined, there are no transaction costs (i.e., the costs of, or impediments to, the transfer of property rights), and efficiency is defined with no reference to the distribution of income among parties. This deeply ambivalent general result, known as the Coase theorem, implies that harmful interference can be controlled by acting on transmitted power - as traditionally occurred in the 'command-and-control' regime of spectrum management, or on aggregate received power.

Accordingly, the SUR concept, illustrated in Fig. 4, is based on defining a certain *aggregate received power* threshold as statistically averaged value throughout the SUR licence area. The hard limit would apply at the edge of the licence area, which could be a country border or some administrative delineation for regional SUR licences. An additional, lower, SUR limit would be set to control average interference power outside the band assigned to the licence holder, i.e., in effect limiting the adjacent band interference. Typically, the SUR thresholds could be expressed in Power-Flux Density units.

It is important to make a distinction between the SUR and the recently popular Block Edge Mask (BEM) concept. The SUR defines the in-band and adjacent-band limits on the received power as aggregated and averaged from all stations operated in the licence area, whereas the BEM defines the limits on transmit power of individual transmitters. In other words, the BEM effectively locks in the presumption of certain type of transmitters to be used in given band, and even then, the evaluation of interference potential would require knowing where those transmitters are located, their density. The SUR gives licence holder the full freedom to choose the type of deployed transmitters while providing certainty to neighbouring users (both geographically and in spectrum domain) with regards to the level of expected interference. Thus, the SUR approach to regulating spectrum access gives licence holders full freedom to decide themselves on the kind of spectrum use, services, and applications.

Ofcom used SUR for an auction of L-band (1400 MHz) spectrum in 2006. The spectrum was initially awarded to Qualcomm, who originally bought it for mobile TV application; then Qualcomm sold the spectrum on to Vodafone who repurposed it for 4G. The SUR approach is thus proven and can be adopted by any regulator in suitable bands.

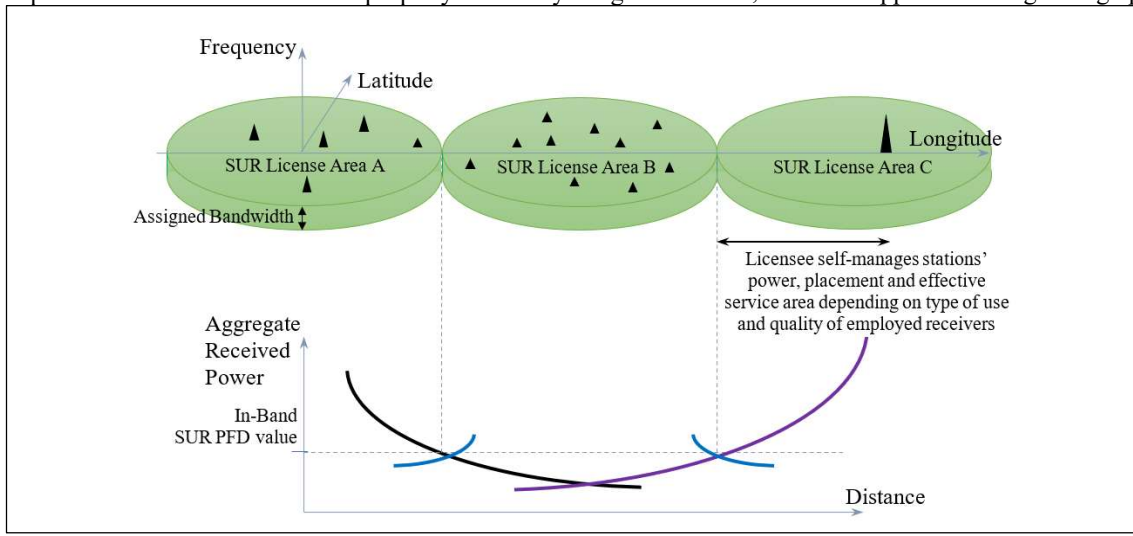

Beyond the inherent flexibility of spectrum utilisation, the SUR-based licence area definition has the additional advantage of going some way to resolving the "receiver problem" [13]. It manifests when users deploy sub-standard receivers when interference is low in neighbouring bands, and then complain if interference grows. An SUR sets out clearly the level of interference into neighbouring bands and hence those designing equipment for those bands should take this into account. If they do not, then the fault is clearly theirs since they designed a receiver unable to tolerate potential interference. For "too big to fail" services, regulators might want to test receivers to make sure that they can accommodate the interference allowed by the SURs.

Important to note, that the regulators should not be obsessing about getting the delineation of SUR border conditions perfect. What is more important, is to make the limits clear and standardised, associate them with economically efficient flexible-use spectrum bundles without excessive fragmentation, and then deliver them to responsible economic actors and let the market do the rest [12]. Having an explicit interference limit would also establish the necessary basis for market players to carry out any bilateral bargaining regarding spectrum access and interference avoidance [13].

### B. Full shared access

While service neutrality, as afforded by SUR or similar flexible spectrum usage rights, can improve the utilisation of the exclusively licensed spectrum, another complementary approach is to allow shared use of spectrum bands by multiple users from the same or different user groups.

Some NRAs already allow managed shared access in certain bands [10]. But this is still very much the exception rather than the rule. There is no reason why managed shared access cannot be extended to multiple bands, albeit the conditions for shared access will vary significantly from band to band.

For example, the 6 GHz band, recently opened for mobile access, may be an excellent candidate for shared use by unlicensed Wi-Fi as well as cellular networks [14].

It may be worth mentioning in this regard that the recent spectrum management innovations, such as those implemented in database-driven DSA systems, go a long way in providing shared spectrum access platforms that improve the spectrum access and use efficiency [10]. The two notable examples of well-developed DSA concepts are the Citizens Broadband Radio Service (CBRS) used in the 3.5 GHz band in the US, and Licensed Shared Access (LSA) in Europe [15].

Looking forward, a recent DARPA program provided practical proof of concept that decentralized spectrum management based on DSA paradigm, and assisted by Artificial Intelligence/Machine Learning (AI/ML) solutions, could lead to more efficient utilisation of spectrum bands than is achievable with centralized spectrum assignment (cf. App. C in [9]).

Such innovative DSA solutions enable various levels of both exclusive and opportunistic spectrum access and plethora of innovative uses. Even more importantly, such systems constitute a crucial first practical step towards making spectrum more commodity-like, a transformation that would open way for true flexibility, ubiquitous spectrum sharing and trading.

### C. Full trading

Many regulators already allow trading. However, none allow changing type of spectrum use through trading. Hence, the trades that would generate the most value – those that take spectrum from a low-value usage to a high-value usage – cannot occur.

Furthermore, there is a number of other factors that tend to reduce the occurrence and efficiency of spectrum trades [11]. For instance, selling spectrum to a competitor may lead to increased competition and resulting loss of seller's market power. This will have an effect of inflating the price of spectrum to the point where it is perceived to compensate sufficiently for potential harm of increased competition. That threshold rises proportionally with the growth in market concentration. This and similar factors resulted in little actual trading in the two decades since trading started to be introduced.

Also, given that the regulators kept at diligently clearing and auctioning ever new bands for mobile use, this has removed much incentive for Mobile Network Operators (MNOs) to engage in spectrum trading or DSA schemes. A case in point is the observed reluctance of European MNOs to support implementing the standardised LSA solution after they received new trenches of exclusive 5G spectrum in 700 MHz and 3.5 GHz bands [15].

The introduction of SUR, along with the licensing of the "last" potential bands in the prime mobile spectrum below 6 GHz, would resolve most of these impediments, allowing the existing mechanisms for trading to work better. Hence, little additional change is needed here other than in some cases easing the requirement for pre-approval of trades to reduce bureaucratic burden.

It should be however mentioned that there needs to be distinction made between "wholesale" and "retail" trading scenarios. It may be one thing to procure a band with nation-wide usage rights (say broadcasting) and change it to something else (say cellular mobile). But in situation where a band is littered with myriad of small users, that would call for some different approach [9].

### D. Minimising Regulatory Intervention

Given the unavoidable occurrences of interference, the NRAs will need to intervene from time to time [10]. The question is however to what degree the government needs to get involved [5], [7]. It would help the spectrum market if the reasons for intervention and the process of intervention were as clear and predictable as possible. This could be done by having some clearly, even if not necessarily exhaustively, defined property rights, such as SURs, delineating the borderline conditions between adjacent spectrum users.

Otherwise, the intervention can often become adversarial. For example, in 3.4-4.2 GHz there is debate as to whether the band should all be auctioned to MNOs or whether there should be set-asides or sharing options for industry verticals and other non-MNO use. Some satellite users are concerned about interference into their systems.

The final decision rests with NRAs, although often informed by regional and global decisions such as those taken at the WRCs. Naturally, those with an interest lobby the regulators strongly. This almost invariably means over-stating their case for the services that could be delivered if only they had access to the band. But the incumbents respond by also over-stating the impact that interference will have on them. Competing parties end up bickering and disputing each other's claims. Regulators do their best, gathering input through consultations, while preparing for legal action from losing parties. This is a slow and expensive process, with significant risk of regulatory capture.

A better way is to use unbiased forecasts that were not inflated by protagonists. This can be achieved by NRAs establishing independent national spectrum advisory panels, comprised of subject matter experts that have high reputation in the industry and proven track record of specialist projects, publications and similar past activities within the radiocommunications development field.

In parallel it makes sense to build in more flexibility to allow allocation decisions to be changed as the future unfolds. Historically this has been near-impossible, e.g., once an unlicensed band has users in it, finding them and removing them is very difficult. But greater use of databases with more conditional usage might increase flexibility. For example, the 6 GHz band access solution could require database recording for all unlicensed and licensed use with greater rights for licensed users – along the lines of the CBRS concept. If licensed usage proved to be the best outcome it would naturally crowd out unlicensed use. If not, unlicensed would flourish.

This approach to cases where intervention is needed would be faster, less adversarial, and more evidence-based than the current, very unsatisfactory, method.

## V. The Path to Improvement

As of today, service neutral licensing is not adopted anywhere and managed spectrum sharing like that enabled by DSA platforms is only used in a very few instances. Broadly this is because it is hard to introduce change, regulators are institutionally risk-averse, and many incumbents prefer a status-quo they know rather than a less-certain future.

SURs are more complex than the current regime, and regulators and legacy users have tended not to see sufficient benefit to champion their introduction. In addition, sharing tends to be fought by current band owners who fear interference and generally have nothing to gain.

Even though the international underpinnings of spectrum management may seem to be entrenching long formed status quo, the changes should be possible and may be started by one entrepreneurial country and then spread to others.

The notable examples of such seminal changes and spectrum management innovations, first taken nationally then backed by RR amendments to provide global harmonisation, include introduction of Wi-Fi in 2.4 GHz ISM band promoted first by the US before being picked up internationally. Similarly, Europe led with HIPERLAN initiative in 5 GHz band that relied on use of innovative spectrum access solutions like Dynamic Frequency Selection, the initiative that was then picked up by the US and further worldwide [6].

Regarding the possible migration paths towards a SUR or similar licensing regime, change is, in principle, straightforward albeit it would need a gradual evolution. Important is to make the first step – for example a regulator need only change some mobile or broadcasting spectrum licenses to SUR format to introduce clearly defined tradeable spectrum rights with liberal use conditions. If all regulators begin on this path, the change can gradually happen worldwide. A few pioneering regulators, willing to push the boundaries, are the most likely catalyst for change.

Further push for spectrum access liberalisation and related innovation could be provided by global IT platform owners – such as Amazon and Alphabet. They have recently started entering wireless communication service markets by offering corporate customers "one-stop-shop" solutions to implement local wireless networks known as "Private 5G". Since such platform owners do not have any of their own exclusive spectrum holdings, nor may expect each customer to engage with regulator to obtain one themselves, using shared spectrum access offers very attractive business development path.

## VI. Conclusions and a Vision for the Future

It is important to stress that the reforms and approaches discussed here do not need to be applied in a "big bang" manner across the entire spectrum. Quite to the contrary, they might be introduced gradually whenever the new band is being opened for commercial licence awards. At the end it might be relevant to managing just a portion of overall spectrum that is used for providing commercial and consumer-oriented wireless services, such as land mobile and broadcasting services, where the innovation and evolutionary change of use would be of most relevance. Below we provide one such vision of how the proposed approach may play out in practice.

In country X the use of UHF broadcasting is steadily decreasing. A new entrant has an innovative idea of using some of the spectrum to deliver a new IoT solution, providing excellent coverage and much higher capacity and flexibility than other solutions. The IoT operator approaches the broadcaster and negotiates a trade of some of the spectrum. It then re-purposes the spectrum from broadcasting to mobile and deploys a network all without any regulatory involvement.

As a result, the technology is deployed five years sooner than in other countries, giving country X a clear competitive advantage, encouraging more innovation and significantly growing the contribution spectrum makes to GDP. At the same time, country X spends far less on its regulator than other similar countries since its involvement is limited to those areas where socially desirable outcomes are determined to be appropriate.

While this vision may be simplified and idealistic, it should be possible to guide the NRAs and other stakeholders towards that vision by moving step by step towards truly flexible and liberal spectrum management.

The research community may help along this way by developing new case studies and proposing practical solutions of spectrum sharing and coexistence for the evolving radiocommunication systems and applications, such as 6G and Non-Terrestrial Networks. The use of AI/ML deserves particular attention as novel enabling factors to aid the migration from centralized to decentralized spectrum management.

## ACKNOWLEDGEMENT

The authors would like to thank the anonymous reviewers for many valuable comments and suggestions that helped to improve this paper.

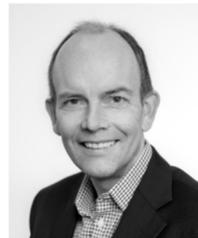

**William Webb** (Fellow, IEEE) holds PhD, MBA and DSc degrees in electronic engineering from Southampton University, UK and multiple other honorary degrees.

He is CTO at Access Partnership. He was one of the founding directors of Neul, a company developing machine-to-machine technologies. Prior to this William was a Director at Ofcom where he managed a team providing technical advice and performing research. He was IET President 2014-2015.

Professor Webb has published 17 books, 100 papers, and 18 patents. He is a Fellow of the Royal Academy of Engineering, the IEEE and the IET and a Visiting Professor at Southampton University. In 2018 he was awarded the IET's Mountbatten Medal for technology entrepreneurship.

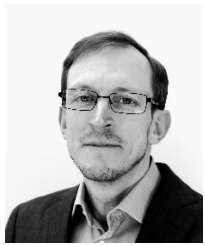

**Arturas Medeisis** (Senior Member, IEEE) holds PhD in telecommunications engineering from Kaunas University of Technology, Lithuania.

He is Adjunct Professor at VILNIUS TECH and Chief Scientific Officer at Cellular Expert UAB, Vilnius, Lithuania. Arturas worked in the field of radio spectrum management for 28 years, including positions with the National Regulatory Authority of Lithuania, European Radiocommunication Office and the Telecommunications Development Bureau of the ITU. He has edited two books and authored or co-authored numerous papers and book chapters on spectrum management and development of radiocommunication technologies.

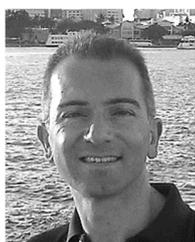

**Leo Fulvio Minervini** holds PhD from the University of Warwick and LL.M (Law and Economics) from the University of Manchester, UK.

He is senior researcher with tenure track of Public Finance at the University of Macerata, Italy. His major research interests include the economics of public goods, the regulation of public utilities, and the political economy approach. He has published works mostly on radio spectrum management, independent regulatory authorities, and public expenditure. He has been involved in a number of national and international projects on regulatory matters.